\documentclass{ws-procs9x6}
\usepackage{amsmath}
\usepackage{amssymb}
\usepackage{bbold}

\newcommand{\eqn}[1]{(\ref{#1})}  
\newcommand{\ospsixtwo}{\text{osp}^*(8|4)}

\begin{document}
\title{Many faces of D-branes: from flat space, via AdS to
  pp-waves$^{*}$}
\makeatother
\author{Marija Zamaklar}
\address{MPI f\"ur Gravitationsphysik\\
Am M\"uhlenberg 1\\
14476 Golm, GERMANY}
\maketitle
\footnotetext{${}^*$~To appear in the proceedings of the Balkan
  Workshop 2003, 29 Aug.~-~2~Sept., Vrnja\v{c}ka Banja, Serbia.}

\abstracts{We review recent studies of branes in $\text{AdS}\times S$ and
pp-wave spaces using effective action methods (probe branes
and supergravity). We also summarise results on an algebraic study of
D-branes in these spaces, using extensions of the superisometry
algebras which include brane charges.}

\section{Introduction}

Since the discovery in 1995, D-branes have become a center of
intensive research of the string theory community. In this process a
lot has been learnt about various manifestations of these objects.  It
has become clear that, depending on the regime in which one works,
D-branes can be described using a variety of techniques. In situations
with a small number of branes and weak string coupling, methods of
open string conformal field theory are appropriate. 
These techiques however, have been applied mainly to the study of
D-branes in very restricted classes of backgrounds, the main obstacle
beeing one's inability to quantise strings in arbitrary backgrounds.

This problem can be partially circumvented by restricting one's
interest to low-energy processes in the target space and the
worldvolume of branes.  In this case a description of branes using
effective actions becomes viable. The low-energy dynamics of the
\emph{closed} strings is governed by various supergravity actions
(which are just various supersymmetric generalisations of the
Einstein-Hilbert action).  The effective action for \emph{open}
strings is the Dirac-Born-Infeld action, which (as the name suggests)
is a generalisation of the Dirac action for the relativistic membrane
to higher dimensional objects, modified to include the brane's
``inner'' degrees of freedom, i.e.~gauge fields. The latter are
included in a fashion proposed a long time ago by Born and Infeld in
order to regularise the infinite electromagnetic energy of a classical
electron. The total (bulk plus brane) effective action is quite
complicated due to the non-trivial coupling between the brane and
supergravity fields. Hence, in order to use this action, one is often
forced to simplify the problem further.  Increasing the number of
branes, for example, leads to the regime where gravitational
back-reaction of the branes cannot be neglected, while the gauge
theory on the worldvolume of the branes becomes strongly coupled. In
this regime branes can be described as purely gravitational solutions,
using only the bulk effective action. On the other side, when the
number of branes is very small, the \emph{probe brane} approach is
appropriate: only the worldvolume excitations (i.e.~scalars and gauge
fields) are dynamical fields, while the background is ``frozen''.  In
the first two parts of this report we will partially survey our recent
study of D-branes in AdS and plane-wave (pp-wave) geometries using the
supergravity and Dirac-Born-Infeld effective actions.

Finally, a very powerful method for classifying possible brane
configurations in arbitrary backgrounds is the so-called
\emph{algebraic method}. The full information about the
\emph{non-perturbative} spectrum of string theory (in a given
background) is encoded in the ``central'' extensions of the
appropriate superisometry algebra of the background. Unfortunately,
the explicit forms of the algebras are essentially not known beyond
flat space. Recently, however, we have made an important step in
understanding the construction of central extensions of AdS and
pp-wave superisometry algebras.  In the last part of this survey we
report on these results.

\section{The probe brane approach}
\subsection{From flat space to AdS}
\label{fl-Ad}
The problem of understanding the full brane-background system is
simplified dramatically by making a restriction to the effective
actions, and by further restricting to the probe brane
approach. However, this still leaves one with a generically
complicated action which has to be solved in order to find the exact
embedding of the brane surface in the target space. In generic
backgrounds, and in cases of supersymmetric brane configurations, one
often uses kappa-symmetry or calibration methods in order to replace
the second-order differential equations with first order, BPS-like
equations. In practice however, both of these methods are ad hoc,
since they both require good intuition about the ansatze for the
brane embeddings.

In special backgrounds, such as AdS or pp-waves, the situation is
simpler: \emph{supersymmetric} brane configurations can be ``derived''
from the configurations of branes in flat space.  Namely, given a
brane configuration in flat space one first replaces some of the
branes in the configuration by their supergravity solutions and
subsequently focuses on the corresponding near-horizon geometries,
while keeping the remaining branes as probes.  The actual equations
for the embedding of the probe brane can usually be deduced directly
from the flat space equations, using Poincar\'e coordinates for the
AdS background. This is due to the fact that in this coordinate
system, the relation of AdS coordinates to flat space Cartesian
coordinates is direct. Moreover, it turns out that the equations that
describe the embedding of the brane in flat space also describe the
solutions of the \emph{DBI equations of motion} in the near horizon
geometry in Poincar\'e coordinates. The essential reason why this
inheritance property holds is that the brane configurations in
question are supersymmetric.  In contrast, if the brane system is
\emph{non-supersymmetric}, this logic does not hold.  For example,
consider a circular $D1$~string in the space transverse to a
$D3$~brane. In flat space one can easily derive the solution that
describes shrinking of the string, and then one can show that this
solution (in Cartesian coordinates), when interpreted in Poincar\'e
coordinates, \emph{does not} solve the DBI equations of motion of the
$D1$~string in the $\text{AdS}_5 \times S^5$ space.

Instead of listing all possible brane configurations which one can
derive using this method, lets us illustrate it on a very simple
example of two $D3$~branes intersecting over a string,
\begin{equation}
\label{conf1}
\begin{array}{lcccccccccc}
D3   & 0 & 1 & 2 & 3 & - & - & - & - & - & -  \\ 
D3   & 0 & 1 & - & - & 4& 5 & - & - & - & -  \, .\\ 
\end{array}
\end{equation}
We take  the first brane to create the background, while the second
brane is treated as a probe.  In flat space the embedding equations of
the second brane are given by
\begin{equation}
\label{eq:embe}
x_i = c_i=\text{const}. \, , \quad (i=2,6,7,8,9) \, .
\end{equation}
Next, we replace the first brane with the near horison limit of its
supergravity solution (~i.e. the $\text{AdS}_5\times S^5$ space),
which in Poincar\'e coordinates yields
\begin{equation}
\label{ads}
\begin{aligned}
{\rm d}s^2  &= R^2 u^2 \big(-{\rm d}t^2 + {\rm d}x_1^2 + {\rm d}x_2^2 + {\rm d}x_3^2 \big) +
 {R^2 \over u^2} \Big( {\rm d}x_4^2 + \ldots + {\rm d}x_9^2
\Big)\,,  \\[1ex]
u^2 &= x_4^2 + ... +x_9^2 \quad\Rightarrow\quad {\rm d}x_4^2 +\ldots+ {\rm
 d}x_9^2 = {\rm d}u^2 + u^2 {\rm d}\Omega_5^2  \, .
\end{aligned}
\end{equation}
When all coefficients $c_i$ in (\ref{eq:embe}) are zero, we see that
the equations (\ref{eq:embe}) define a maximal-curvature $\text{AdS}_2
\times S^1$ submanifold of
(\ref{ads})~\cite{Bilal:1998ck,Gutowski:1999tu}. In the case when some
of the $c_i \neq 0$, the equations~(\ref{eq:embe}) solve the $D3$ DBI
action in the (full) $D3$~brane supergravity background. However, when
taking the near horizon limit, one additionally needs to scale the
parameters~$c_i$ to zero in order for the solution to survive this
limit. As we focus on the region near the $D3$~brane that becomes the
background, we simultaneously have to bring the probe $D3$~brane
closer and closer to the horizon. The resulting geometry of the
$D3$~brane probe describes a brane which starts at the AdS boundary,
extends in the $u$-direction up to some point and then folds back to
the boundary~\cite{Karch:2002sh}.

Recently,~\cite{Sarkissian:2003jn} we have extended this analysis to
the cases of supersymmetric brane configuration intersecting under an
\emph{angle} in flat space. As in the previous situation, the
inheritance property goes through due to supersymmetry. However, the
resulting brane geometries are qualitatively different from the one
previously discussed, since the branes now non-trivially mix the AdS
and sphere submanifolds: the worldvolume surfaces are not factorisable
into a product of AdS and sphere submanifolds.

Another new type of brane that has appeared
in~\cite{Sarkissian:2003jn} is a brane with mixed worldvolume flux
(i.e.~where the worldvolume two-form has one index in the sphere part
and one index in the AdS part). This brane wraps
\emph{non-supersymmetric} target space cycles and is stabilised only
after the mixed worldvolume flux is turned on.  To construct this
brane, one starts from the flat space configuration of branes
intersecting under an angle and performs a T-duality transformation in
such a way that the brane which will be replaced with the background
\emph{does not} carry any worldvolume flux, while the brane which will
become a probe carries flux. Then, as before, one takes the near
horizon limit of this configuration.

\subsection{From AdS to pp-waves}
\label{Ad-pp}
It was realised a long time ago by Penrose that an infinitely boosted
observer in an \emph{arbitrary} spacetime, in a neighborhood of its
geodesic, observes a very simplified background geometry: the geometry
of a gravitational wave. This dramatic simplification has recently
been used extensively for a direct check of the gauge-gravity
(AdS/CFT) correspondence~\cite{Berenstein:2002jq}, avoiding the
standard strong-weak coupling problems.

On the gravity side, the Penrose limit amounts to a suitable rescaling
of the coordinates and parameters characterising the (super)gravity
solution, in such a way that one focuses on the region close to an
arbitrary null geodesic.  In the same way in which the background
undergoes simplification, so do different objects present in the
initial space. The geometry of the resulting branes can easily be
derived by rescaling the embedding equations of the branes in the same
way as the target space coordinates. Since the pp-wave space is
\emph{homogeneous} but \emph{not isotropic}, there are three basic
families of D-branes which appear in the limit, depending on the
relative orientation of the brane and the
wave:~\cite{Skenderis:2002vf} \emph{longitudinal D-branes} for which
the pp-wave propagates along the worldvolume of the D-brane,
\emph{transversal D-branes} for which the pp-wave propagates in a
direction transverse to the D-brane but the timelike direction is
along its worldvolume, and \emph{instantonic D-branes} for which both
the direction in which the pp-wave propagates and the timelike
direction are transverse to the D-brane.  The first class of branes
originates from AdS branes for which the geodesic along which the
boost was performed belongs to the worldvolume of the brane (before
the limit). For the second class, the brane is co-moving with
the observer along the the geodesic (i.e.~it was infinitely
boosted). The third class of branes can be obtained from the first
class by a formal T-duality in the timelike direction of the wave.

Based on isometries, the pp-wave coordinates can be split into three
groups: the ``lightcone coordinates'' $u$ and $v$, and two
four-dimensional subspaces with $SO(4)\times SO(4)$ isometry
group. The split of the transverse coordinates is due to the
nonvanishing 5-form flux. In the case of longitudinal branes, the
worldvolume coordinates split accordingly into three sets: the
``lightcone coordinates'' $u$ and $v$, $m$ coordinates along the first
$SO(4)$ subspace and $n$ along the second $SO(4)$. A $Dp$~brane
($m+n=p-1$) with such orientation is denoted with $(+,-,m,n)$.  The
number of preserved supersymmetries depends on the values for
$(n,m)$:\cite{Skenderis:2002vf}
\begin{itemize}
\item $1/2$-BPS D-branes with embedding $(+,-,m+2,m)$,\\ for  $m=1,\ldots,4$,
\item $1/4$-BPS D-string with embedding $(+,-,0,0)$,
\item non-supersymmetric D-branes with embedding $(+,-,m,m)$,\\
for $m=1,2,3$.
\end{itemize}
All these results are valid for the brane placed at the ``origin'' of
the pp-wave. If we \emph{rigidly} move the first or second type of brane
outside the origin~(without turning worldvolume fluxes), supersymmetry
is always reduced to $1/4$.

However, the previous three classes do not capture all the branes
which can appear in
pp-waves~\cite{Sarkissian:2003jn,Sarkissian:2003yw}.  In the process
of Penrose rescaling, not all objects of the initial space will be
inherited by the final wave geometry.  It is usually said that in
order to have a nontrivial Penrose limit of a brane in some
background, one needs to take the limit along a geodesic which belongs
to the brane. This statement is intuitively understandable: in the
Penrose limit an infinitesimal region around the geodesic gets zoomed
out. Hence, those parts of the brane which are placed at some nonzero
distance from the geodesic get pushed off to infinity.  However, this
reasoning can be circumvented if the distance between the geodesic and
the brane is determined by free parameters of the
solution~\cite{Sarkissian:2003jn}. In that case one can take the
Penrose limit along a geodesic that \emph{does not} belong to the
brane, as long as the parameter labeling the brane in a family of
solutions is appropriately scaled.

For example, let us consider the family of solutions corresponding to
two intersecting D-branes and let us take the Penrose geodesic to lie
on the first of the two branes. Then the Penrose limit of the second
brane can be nontrivial if, while taking the Penrose limit of the
target space metric, we simultaneously scale the angle between the two
branes to zero.  It should be emphasized that the final configuration
obtained in this way is different from the one which is obtained by
first sending the angle to zero and then taking the Penrose limit of
the metric. Namely, if we first set the angle to zero, the problem
will reduce to taking the limit along a geodesic that belongs to the
worldvolume of the brane, which has been discussed already.  However,
if we follow the procedure outlined above, the Penrose limit of the
second D-brane is a brane with a relativistic pulse propagating on its
worldvolume (i.e.~with some of $x_i=\text{const}.$ getting replaced by
$x_i(x^+)$).  The precise form of this worldvolume wave carries
information about the position of the brane with respect to the
geodesic before the limit was taken~\cite{Sarkissian:2003jn}.

\section{The supergravity approach}

When the number of branes in some space becomes very large, the probe
brane approach is inadequate and a supergravity description takes
over.  Finding supergravity solutions for D-branes in AdS spaces is,
however, still essentially an open problem; the explicit constructions
have been carried out only in a few specific cases.  One of the
reasons for this is that fully localised supergravity solutions for
D-brane intersections in \emph{flat space} are generically not known.
Hence in order to construct the brane solutions in asymptotically AdS
and pp-wave spaces, one has to start from scratch. We will present
here the construction of the (extremal) D-brane solutions in
asymptotically pp-wave spaces.

The main difficulty in constructing these solutions consists of
identifying a coordinate system where the description of the D-brane
is the simplest. This is similar to the problem that one would face if
one would only know about Minkowski space in spherical coordinates and
would try to describe flat D-branes in these coordinates. Cartesian
coordinates are the natural coordinates to describe infinitely
extending D-branes in flat space. So the question that one should
first ask is ``what are the analogues of the Cartesian coordinates for
D-branes in pp-wave backgrounds?''.  The answer to this question is
more complicated than in flat space, as it depends very much on what
kind of D-branes one considers. It was
shown~\cite{Bain:2002nq,Bain:2002tq} that Brinkman coordinates are the
natural coordinates for a description of $1/4$-BPS and
nonsupersymmetric D-branes, while the natural coordinates for the
$1/2$-BPS D-branes are the Rosen coordinates.

For the metric part of the ansatz, one writes a simple standard metric
for a superposition of D-branes with waves,
\begin{multline}
\label{ansatza}
{\rm d}s^2 = H(y,y')^{-{1\over 2}} \Big(
    2 {\rm d}u\big({\rm d}v + S(x,x',y,y'){\rm d}u\big) - 
    {\rm d}\vec{x}^{\,2} - {\rm d}\vec{x}'^{\,2} \Big) \\
  - H(y,y')^{1\over 2} ({\rm d}\vec{y}^{\,2} + {\rm d}\vec{y}'^{\,2})\, .
\end{multline}
The metric is given in the string frame, and the D-brane worldvolume
coordinates are ($u,v, x^i = x^1,\ldots, x^m, x'^{I}=
x'^1,\ldots,x'^n$), while the directions transverse to the D-brane are
$(y^a = y^1,\ldots,y^{(4-m)}, y'^A = y'^1,\ldots,y'^{(4-n)})$.
The function~$H$ characterising the D-brane is at this stage
allowed to depend on all transverse coordinates~\mbox{$y$, $y'$}.
The ansatz for the RR field strength and the dilaton reads
\begin{align}
\label{ansatzb}
F_{[p+1]} &= {\rm d}u \wedge {\rm d}v \wedge {\rm d}x^1 \wedge \cdots
 \wedge {\rm d}x^m \wedge {\rm d}x'^1 \cdots \wedge {\rm d}x'^n \wedge
 {\rm d}H^{-1} \, , \\
\label{ansatzc}
F^{\rm w}_{[5]} &= F_{[5]}^{(1)}  + * F_{[5]}^{(1)} \, , \nonumber \\
F_{[5]}^{(1)} =& W(z^{\mu}) \, {\rm d}u\wedge\left(  
  {\rm d}x^1 \wedge \cdots \wedge {\rm d}x^m \wedge {\rm d}y^1 \wedge
\cdots \wedge {\rm d}y^{(4-m)} \right) \, , \\
\label{ansatzd}
e^{\phi} &= H^{3-p \over 4}   \, ,
\end{align}
where `$*$' in $F_{[5]}$ denotes Hodge duality with respect to the
metric~(\ref{ansatza}) and $W(z)$ is an undetermined function which
can depend on all directions transverse to the pp-wave.  In the
case of the $D3$ brane, one also has to add to the form~(\ref{ansatzb}) its
Hodge dual.

One of the main characteristics and perhaps limitations of this ansatz
is that the metric is diagonal in Brinkman coordinates.  This property
forces one to delocalise the \emph{supersymmetric} solutions along
some directions transverse to the brane when solving the equations of
motion.\footnote{This is the same type of restriction that one faces
when constructing supergravity solutions for intersecting D-branes,
with a simple diagonal ansatz.}  The smearing procedure physically
means that one is constructing an array of D-branes of the same type
with an infinitesimally small spacing.  However, as we have seen
before, the probe brane results tell us that, unless we turn on
additional bulk fluxes (sourced by the worldvolume fluxes of the
$1/2$-BPS D-branes), a periodic array of \emph{rigid} D-branes in
\emph{Brinkman} coordinates with orientation $(+,-,n+2,n)$ is only one
quarter supersymmetric.  Hence the supersymmetric solutions that we
find due to the smearing procedure are only $1/4$~BPS.  However, these
restrictions have to be imposed only on the harmonic function
characterising the D-brane, and not on the function characterising the
pp-wave. Hence, all our solutions asymptotically tend to the
\emph{unmodified} Hpp-wave.  Also, despite the simplicity of the
ansatz, the \emph{non-supersymmetric} solutions, describing branes
with $(+,-,m,m)$ orientation, are fully localised.

Plugging the ansatz (\ref{ansatza})-(\ref{ansatzd}) into the equations
of motion and the Bianchi identities, one obtains solutions with the
following characteristics, which depend on the orientation of the
branes.  The presence of the D-brane modifies the function~$S$ which
characterises the pp-wave, while the function~$H$ (which specifies the
D-brane) is completely unmodified by the presence of the
wave. Therefore, this ansatz does not catch the \emph{back-reaction of
the pp-wave on the D-brane}. For a generic embedding of the D-brane,
one expects that the (fully localised) D-brane is modified by the
wave. However, as our fully localised, nonsupersymmetric solution
demonstrates, this does not have to hold for specific embeddings. By
examining the behavior of the radially infalling geodesics, one
discovers that if the pure pp-wave was focusing the geodesics, this
attractive behaviour is strengthened in the presence of a
supersymmetric brane, as one would expect. Surprisingly, however, the
non-supersymmetric geometries exhibit repulsion behavior.

\section{The algebraic approach}

Rather surprisingly, a modification of the superalgebra of
anti-de-Sitter backgrounds which accounts for the presence of D-branes
in the string spectrum is still unknown. At an algebraic level,
D-branes manifest themselves through non-zero expectation values of
bosonic tensorial charges. There exists a widespread, but incorrect,
belief that the inclusion of these brane charges into the
anti-de-Sitter superalgebras follows the well-known flat-space
pattern. In flat space, the inclusion of brane charges leads to a
rather minimal modification of the super-Poincar\'e algebra: the
bosonic tensorial charges appear on the right-hand side of the
anti-commutator of supercharges, transform as tensors under the
Lorentz boosts and rotations, while they commute with all other
generators.  The brane charges are therefore often loosely called
\mbox{``central''}, and the resulting algebra is referred to as the
maximal bosonic \mbox{``central''} extension of the super-Poincar\'e
algebra. However, despite several attempts to construct a similar
modification of anti-de-Sitter superalgebras, a \emph{physically
satisfactory} solution is as of yet unknown.

There are two basic physical requirements which have to be satisfied
by an anti-de-Sitter algebra which is modified to include brane
charges. The algebra has to include at least the brane charges which
correspond to all D-branes that are already known to exist, and it
also has to admit at least the supergraviton multiplet in its
spectrum.  Mathematically consistent modifications of anti-de-Sitter
superalgebras can be constructed, but all existing proposals fail to
satisfy one or both of these physical criteria.~\cite{Meessen:2003yi} 
In \cite{Peeters:2003vz} we have identified 
a simple reason why previous attempts to extend
anti-de-Sitter superalgebras with brane charges have failed: such
extensions are \emph{only} physically acceptable when one adds new
\emph{fermionic} brane charges as well.

The necessity of including new fermionic brane charges into the
modified algebra can be understood from a very simple argument based
on Jacobi identities, in combination with the two physical
requirements just mentioned \cite{Peeters:2003vz}. Consider an
anti-de-Sitter superisometry algebra, or a pp-wave contraction of
it. The bracket of supercharges can, very symbolically, be written in
the form
\begin{equation}
\{ Q_\alpha, Q_\beta \} = (\Gamma^{AB})_{\alpha\beta} M_{AB}\, ,
\end{equation}
where $Q$ and $M$ are the supercharges and rotation generators
respectively (we have grouped together momentum and rotation
generators by using a notation in the embedding space). Suppose now
that we add a bosonic tensorial brane charge~$Z$ on the right-hand
side of this bracket. This extension has to be made consistently with
the Jacobi identities. Consider the $(Q,Q,Z)$ identity, which takes
the symbolic form
\begin{equation}
\label{e:QQZ}
\begin{aligned}
( Q_\alpha, Q_\beta, Z ) 
&=  [\{Q_\alpha,Q_\beta\}, Z] 
- \big\{[Q_\alpha, Z], Q_\beta\big\}
- \big\{[Q_\beta, Z], Q_\alpha\big\}  \\[1ex]
&= (\Gamma^{AB})_{\alpha\beta} [ M_{AB}, Z]
 -2\,\big\{[Q_{(\alpha}, Z],Q_{\beta)}\big\}\,.
\end{aligned}
\end{equation}
As the brane charge~$Z$ is a tensor charge, it will transform
non-trivially under the rotation generators. This implies that the
first term of~\eqn{e:QQZ} will not vanish. The Jacobi identity can
then only hold if~$Z$ \emph{also} transforms non-trivially under the action
of the supersymmetry generators! (In flat space, only the
vanishing bracket $[P,Z]$ appears in the first term of the Jacobi
identity~\eqn{e:QQZ}, because in that case the $\{Q,Q\}$
anti-commutator closes on the translation generators).  The simplest
option is to assume that \emph{no new fermionic charges} should be
introduced, and that therefore symbolically
\begin{equation}
\label{e:QZisQ}
[ Q_\alpha, Z ] = Q_\alpha \, .
\end{equation}
Although it is possible to construct an algebra based on~\eqn{e:QZisQ}
which satisfies all Jacobi identities, it is physically
unsatisfactory~\cite{Meessen:2003yi}. The essential reason is that
brackets like~\eqn{e:QZisQ} are incompatible with multiplets on which
the brane charge is zero (the left-hand side would vanish on all
states in the multiplet, while the right-hand side is not zero).  In
other words, one cannot ``turn off'' the brane charges.
The only other way out is to add \emph{new
fermionic charges}~$Q'_\alpha$ to the algebra, such that~\eqn{e:QZisQ}
is replaced with
\begin{equation}
[ Q_\alpha, Z ] = Q'_\alpha \, .
\end{equation}
In this case it becomes possible to find representations in which both
$Z$ and the new charge $Q'_\alpha$ are realised trivially, as expected
for e.g.~the supergraviton multiplet, while still allowing for
multiplets with non-zero brane charges.

This formal argument based on Jacobi identities may come as a
surprise, and one would perhaps find it more convincing to see new
fermionic brane charges appear in \mbox{\emph{explicit}} models. In
\cite{Peeters:2003vz} we have shown that such charges indeed do
appear. In order to show this, we have analysed the world-volume
superalgebras of the supermatrix model and the supermembrane in a
pp-wave limit of the anti-de-Sitter background. These models exhibit,
in the absence of brane charges, a world-volume version of the
superisometry algebra of the background geometry. When bosonic winding
charges are included, the algebra automatically exhibits fermionic
winding charges as well. Moreover, configurations on which these
charges are non-zero can be found explicitly, or can alternatively be
generated from configurations on which the fermionic winding charges
are zero. On the basis of these results we have briefly discussed a
D-brane extension of the $\ospsixtwo$ superisometry algebra with
bosonic as well as fermionic brane charges, which avoids the problems
with purely bosonic modifications as first observed
in~\cite{Meessen:2003yi}. A partial construction of this algebra has
been carried out in \cite{Lee:2004jx}.

\providecommand{\href}[2]{#2}\begingroup\raggedright\endgroup

\end{document}